\documentclass[aps,prl,twocolumn,superscriptaddress,groupedaddress]{revtex4}
\usepackage{subfig}
\usepackage{graphicx}  % needed for figures
\usepackage{dcolumn}   % needed for some tables
\usepackage{bm}        % for math
\usepackage{amssymb}   % for math
\usepackage{slashed}
\usepackage{graphicx}				% Use pdf, png, jpg, or eps with 	
\usepackage{amsmath}
\usepackage{mathtools}
\usepackage{tikz,pgf}
\usepackage{comment}
\usepackage{asymptote}
\usetikzlibrary{arrows,backgrounds}
\usetikzlibrary{fit,scopes,calc,matrix,positioning,decorations.pathmorphing}
\usepackage[all]{xy}
\usepackage{yfonts}

\newcommand{\ket}[1]{\ensuremath{\left|#1\right\rangle}}

\begin{document}
\title{Axion mass and quantum information}
\author{Andrei T. Patrascu}
\address{ELI-NP, Horia Hulubei National Institute for R\&D in Physics and Nuclear Engineering, 30 Reactorului St, Bucharest-Magurele, 077125, Romania}
\begin{abstract}
I present a novel way of linking the shift in the axion mass due to non-perturbative gravitational effects and the quantum information interpretation of the holographic principle. The main result of this article is a novel way of expressing the shift in the axion mass by means of properties of the tensor network used in the construction of the quantum information interpretation of the holographic principle. Indeed it appears that the axion mass shift can be seen as a geometric property of the holographic bulk entanglement. Moreover, the explicit breaking of the continuous Peccei-Quinn symmetry by gravitational instantons can be regarded as a result of including tensors encoding gauge fields on the holographic bulk tensor network. The origin of the mass of light scalars can therefore be associated to entanglement properties leading to a new connection between mass and quantum information. 
\end{abstract}
\maketitle
Axions play a crucial role in our understanding of various mechanisms both in astrophysics [1] and cosmology [2]. They also appear naturally in string theory and are required by the Peccei Quinn mechanism in solving the strong CP problem [3]. Beyond their astrophysical relevance as a mechanism of cooling for neutron stars [4] and as a candidate for light dark matter, axions are particularly important as their presence allows for wormhole solutions [5]. Such solutions, involving gravitational instantons seen from an Euclidean perspective allow us to analyse non-perturbative gravitational effects by looking into topology changing effects induced by the transition $\mathbb{R}^{3}\rightarrow \mathbb{R}^{3}\times S_{3}$ considering the addition of the $S_{3}$ section of the wormhole throat. 
As pseudo-Goldstone bosons, axions obtain their small mass from explicit breaking of the Peccei Quinn symmetry. I show here that the terms capable of producing such masses appear naturally in the tensor network describing entanglement on wormhole geometries. In this way a previously unexplored connection between the origin of mass and quantum entanglement is presented. Beyond its fundamental relevance as a new way of understanding the origin of certain types of masses, such a connection between entanglement and the mass of pseudo-Goldstone bosons presents itself as a new way of understanding high-critical temperature superconductors derived from an approximate SO(5) symmetry as discussed in [12]. In this case too, aside of spontaneous symmetry breaking [13], the explicit breaking of a symmetry must be explained and the mechanisms proposed in the literature lack a sufficiently universal approach. Adding quantum entanglement effects as a source of explicit symmetry breaking and a mechanism for mass production will therefore impact both fundamental high energy physics and effective condensed matter phenomenology. 
On the holographic side, recent studies [6] have shown that we can interpret bulk boundary dualities in terms of quantum error correction codes. The information on the boundary represents a physical expression of the logical bulk information which encodes the required state in a robust way, such that accidental erasures of the boundary qubits are compensated by the bulk encoding prescriptions. A particular way of analysing this phenomenon has been provided in ref. [7] by employing holographic tensor networks with the tensors defined as isometric maps between holographic states described by Hilbert spaces within the bulk. Indeed, using this representation it was possible to re-derive the Ryu-Takayanagi formula and to compute various results related to entanglement entropy in a holographic context [8]. However, what is of interest in this paper is the holographic tensor network construction of a wormhole geometry. This can be obtained by introducing black holes in the central part of the holographic tensor network. This implies removing the central tensor and hence obtaining more free bulk indices which then can be connected to the corresponding free bulk indices obtained similarly on another similar region through the central black hole. A wormhole formed in this way explicitly breaks the Peccei Quinn $U(1)$ symmetry on each side while preserving the axion charge conservation globally. It is the goal of this article to relate various properties of the axion mass and charge to properties of such a holographic tensor network. 
Considering the effective axion potential including both the quantum chromodynamics and non-perturbative gravitational contributions, we obtain 
\begin{equation}
V(a)=\Lambda_{QCD}^{4}cos(\frac{\phi}{f_{a}})+(1/L)^{4}e^{-S_{inst}}cos(\frac{\phi}{f_{a}}+\delta_{1})
\end{equation}
where $L$ is the thickness of the wormhole throat, $\phi$ parametrises the angular direction associated to the axion introduced as the pseudo-Nambu-Goldstone boson arising from the spontaneous symmetry breaking of the Peccei-Quinn symmetry at the scale $f_{a}$, and $\delta_{1}$ is a phase due to the mismatch between the gravitational term and the low energy QCD term. $S_{inst}$ is the instanton action arising from non-perturbative gravitational effects. Below the scale $f_{a}$, the axion obeys a non-linearly realised global $U(1)$ symmetry $\phi\rightarrow \phi+\alpha\cdot f_{a}$ with $\alpha$ being an arbitrary parameter describing a field space shift. Such a symmetry has an associated Noether current $J^{\mu}=f_{a}\partial^{\mu}\phi$ and a charge $Q$ which is the generator of the symmetry. There exists a subgroup of the symmetry group associated to the transformation above, namely the one defined for the discrete shift $\phi\rightarrow \phi+2k\pi f_{a}$, $k\in\mathbb{Z}$. This subgroup represents a gauge symmetry of the system as it is inherent to the angular definition of the axion. The pseudo-scalar axion field can be described starting with the bulk action of a three form $H$ coupled to Einstein gravity 
\begin{equation}
S_{E}=\int d^{4}x\sqrt{g}(-\frac{M_{Pl}}{16\pi}\mathcal{R}+\frac{\mathcal{F}}{2}H_{\mu\nu\rho}H^{\mu\nu\rho})
\end{equation}
where of course the axion field appears from $H_{\mu\nu\rho}=f_{a}^{2}\epsilon_{\mu\nu\rho\sigma}(\partial^{\sigma}\theta)$. As stated before, the coupling of the axion field allows for wormhole solutions where each side of the wormhole throat (each end of a half-wormhole) is associated to an axion charge $+n$ respectively $-n$. This allows us to interpret the whole wormhole as a combination of an instanton with axion charge $+n$ and an anti-instanton with an axion charge $-n$. The gravitational shift of the axion mass has been calculated in [9] and was found to be 
\begin{equation}
m_{a}^{2}\simeq \frac{\Lambda_{QCD}^{4}}{f_{a}^{2}}+\frac{(1/L)^{4}}{f_{a}^{2}}e^{-S_{inst}}
\end{equation}
As the scale $f_{a}$ increases the QCD axion mass $\frac{\Lambda_{QCD}^{4}}{f_{a}^{2}}$ decreases while the gravitational contribution increases becoming more important than the former. This leads to a lower bound of the axion mass and a region where the non-perturbative gravitational corrections dominate. 
Wormhole geometry plays a crucial role in the understanding of axion physics, particularly because it has the capability of establishing the axion mass resulting from the explicit breaking of the Peccei-Quinn symmetry. Half-wormholes induce quantum tunnelling between vacuum states with different axion charges. The observer on one side of the wormhole will see axion charge non-conservation and an explicit breaking of the Peccei Quinn symmetry down to a discrete gauge symmetry. 

Fascinatingly enough, this phenomenon can be described in terms of tensor networks covering the wormhole spacetime geometry. Indeed, by introducing a tensor network over the spacetime with a wormhole geometry in the bulk, it is possible to establish a dictionary between quantum information (entanglement) concepts on one side and explicit symmetry breaking due to wormhole instantons on the other side. In this article I show the first item in this dictionary, namely the connection between entanglement in bulk spacetimes modified by wormhole geometry and explicit symmetry breaking by non-perturbative gravitational effects as an origin for axion mass in the Peccei Quinn theory for the axion. Indeed the axion can be regarded as a pseudo-Goldstone boson associated to the spontaneous breaking of the Peccei-Quinn symmetry. Due to explicit breaking of this symmetry also originating in the context of non-perturbative quantum gravity, we have a small mass associated to the axion. Can such a symmetry breaking effect be explainable through phenomena linked to the quantum information interpretations of the holographic principle? It appears that this is the case. In the standard model, non-linearly realised Abelian global symmetries can be rewritten as local shift symmetries which can be gauged by making use of three form gauge fields [10]. The standard model presents us with anomalous symmetries, Peccei-Quinn being one such case. This can be dualised to local symmetries by means of the Chern-Simons three-forms of the Standard Model gauge group [10]. The strong CP problem then simply becomes the problem of a massless three form field in QCD capable of creating an arbitrary CP-violating constant four-form field in vacuum. The axion appears due to the higgsing of the three-form gauge field and therefore leads to the screening of the four-form field in the vacuum, as has been shown in [10]. Briefly following ref. [10] the strong CP problem in QCD can be written in the three-form language considering a $\theta$-term in $SU(N)$ gauge theory,  with a strong coupling scale $\Lambda$ which can be set to unity, leading to 
\begin{equation}
L=\theta\frac{g^{2}}{32\pi^{2}}F^{a}\tilde{F}^{a}
\end{equation}
where we use the notation $g$ for the gauge coupling and $\tilde{F}^{a}$ for the dual field strength, while $a$ is an $SU(N)$-adjoint index. This term can be written as a four-form field strength 
\begin{equation}
L=\theta \frac{g^{2}}{32\pi^{2}}F^{a}\tilde{F}^{a}=\theta\cdot F=\theta F_{\alpha\beta\gamma\delta}\epsilon^{\alpha\beta\gamma\delta}
\end{equation}
The four-form field strength 
\begin{equation}
F_{\mu\alpha\beta\gamma}=\partial_{[\mu}C_{\alpha\beta\gamma]}
\end{equation}
has been used to define the previous Lagrangian term where the Chern-Simons three form is 
\begin{equation}
C_{\alpha\beta\gamma}=\frac{g^{2}}{8\pi^{2}}Tr[A_{[\alpha}A_{\beta}A_{\gamma]}-\frac{3}{2}A_{[\alpha}\partial_{\beta}A_{\gamma]}]
\end{equation}
and $A_{\alpha}=A^{a}_{\alpha}T^{a}$ is the gauge field matrix and $T^{a}$ are the generators of the gauge group. 
At low energy the three form $C$ becomes a massless field capable of creating a long-range Coulomb type constant force [10]. At low energies therefore the QCD Lagrangian can be written as 
\begin{equation}
L=\theta\cdot F + K(F)+...
\end{equation}
The strong CP-problem in QCD is equivalent to the problem of a constant four-form electric field in the vacuum. The axion solution can be regarded as a higgsing mechanism for this composite three-form. The axion therefore solves the strong-CP problem by giving a gauge invariant mass to the three form field and screening it in the vacuum. Otherwise stated, one can think of promoting $\theta$ to a dynamical field $a$ which transforms the Lagrangian into 
\begin{equation}
L=\frac{f_{a}^{2}}{2}(\partial_{\mu}a)^{2}-a\cdot F -\frac{1}{24}K(F)
\end{equation}
If corrections due to non-perturbative quantum gravity phenomena can take the three form gauge theory out of the Higgs phase and back to the Coulomb phase, the axion solution for the strong CP problem would be in danger. Breaking the global symmetry explicitly by means of quantum gravity is by no means a surprise. Non-perturbative gravitational effects are known not to preserve global symmetries. The connection to entanglement geometry is however new. The three-form field fits into the picture created by a holographic tensor network for the description of entanglement geometry encoded in the bulk spacetime. As has been noted in [11], gauge theories are described by means of a connection and a conjugate electric flux. Therefore, such degrees of freedom can be seen as fitting on the links of a discrete graph model as the one used in the context of holographic quantum networks. Holographic codes have been extended by introducing degrees of freedom on the links of the tensor networks for the pentagon quantum code [11]. Such a code appears as a tilling of a hyperbolic disk with a fundamental unit being a six-indexed tensor $T$. Except for the boundary of this disk, five indices of the tensor (corresponding to legs in the graphic representation of [11]) are connected to adjacent tensors. To every tensor of this type we also have an uncontracted index associated to a bulk degree of freedom. If $T$ is a perfect tensor i.e. describing an isometry from any subset of maximally three indices to the rest, we can have an operator that acts on the bulk, being pushed along any of the remaining legs. Reference [11] shows how to introduce additional bulk degrees of freedom capable of modelling bulk gauge fields by contracting a three-indexed tensor $G$ with a pair of neighbouring bulk input indices originating from two pentagons. Fascinatingly enough, if two such disks are being considered and a central black hole allows the linking to go from one disk to the other we obtain a wormhole spacetime geometry. The linking tensor $G$ will act as a three form field capable of having a role similar to our $C_{\alpha\beta\gamma}$. We can rewrite the axion solution of the strong-CP problem in terms of the antisymmetric field $B_{\mu\nu}$ with the Peccei-Quinn symmetry appearing as a gauge symmetry. The invariant mass term to be added to the Lagrangian is 
\begin{equation}
(\partial_{[\alpha}B_{\beta\gamma]}-C_{\alpha\beta\gamma})^{2}
\end{equation}
Such a mass term screens our QCD four-form field. The three-form Higgs effect means basically just demanding for the axion minimum to be at zero $\theta$. Quantum gravity corrections could jeopardise the Peccei-Quinn solution to the strong CP problem by introducing additional massless three forms $G_{\alpha\beta\gamma}$ capable of contributing to the $B_{\mu\nu}$ mass [10]. This would imply an actual contribution to the axion mass and it would only come from a gauge invariant mass term. In $a$-field terminology, this would amount to mixing of an additional three form. In gravity, the obvious contribution would come from a Chern-Simons three form 
\begin{equation}
G_{\alpha\beta\gamma}=\frac{1}{12}\Gamma_{j\alpha}^{i}\partial_{\beta}\Gamma_{i\gamma}^{j}+\frac{1}{18}\Gamma_{j\alpha}^{i}\Gamma_{k\beta}^{j}\Gamma_{i\gamma}^{k}
\end{equation}
$G$ is expected to give a contribution to the axion mass. But the additional three form $G_{\alpha\beta\gamma}$ linking the central input legs of the holographic tensors insures exactly the connection within the bulk and hence due to the tensors being perfect, we can interpret this as a resulting long-range entanglement in the bulk having as effect the shift in the mass of the axion. The main result therefore is that the holographic tensor network allows us to encode the entanglement within the bulk and it is because of this that we have corrections to the axion mass when quantum gravity effects are being considered. Moreover, when a wormhole solution is present, the central input links allow bulk gauge fields to link through the wormhole throat leading to restoring the explicitly broken symmetry in a global sense. In the language of $B_{\mu\nu}$ fields the field strength 
\begin{equation}
H=\partial_{[\alpha}B_{\beta\gamma]}-C_{\alpha\beta\gamma}-G_{\alpha\beta\gamma}
\end{equation}
will be invariant under both shift gauge symmetries. The contributions of a massless three form field $G$ to the mass of the axion can lead to strong CP violations by means of gravitational effects. One may however avoid such phenomena by restoring a global charge conservation. Indeed, in string theory there are various three forms but also the additional axions that higgs these three forms. Therefore, gravitational instantons are expected to correct the axion mass and couplings without endangering the axion solution to the strong CP problem. This effect can be easily understood from the perspective of holographic tensor networks where such forms act as links between central input indices for polygons in the bulk, connecting them across the bulk, and particularly connecting them across wormholes. 
\\
\begin{figure}
  \includegraphics[width=150pt]{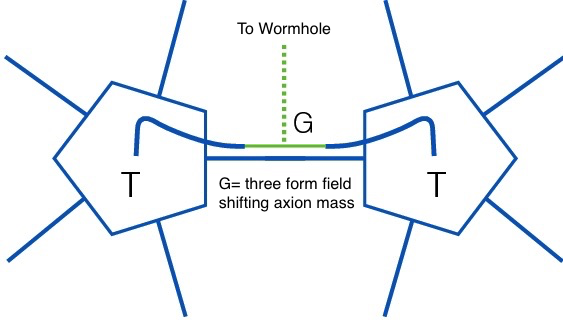}
  \caption{Introducing the three form fields as tensors linking the pentagons of a pentagon holographic code allows us to understand the mass shifting properties of gravitational instantons on axions. Holographic entanglement plays the role in preserving the axion solution to the strong CP problem despite allowing for axion mass shifts. The linking of ref. [11] is reinterpreted in the context of a wormhole instanton.}
  \label{fig:fig1}
\end{figure}
\\
As has been seen in [9], half-wormholes are associated to quantum tunnelling. Given the instanton action $S_{inst}$ for the half-wormhole with axion charge $n$ we can calculate an approximation of the gravitational correction to the axion mass. This can be done by introducing the suitable bulk operators and exploiting the perfect tensors inherent to the holographic tensor network construction.

We obtain, following [9] a low energy effective theory of the form 
\begin{equation}
L=L_{0}[\Phi(x),...]+\sum_{i}\mathcal{O}_{i}[\Phi(x),...]\mathcal{A}_{i}
\end{equation}
where topological wormhole effects are introduced by means of the right hand side sum. As in [9], $\mathcal{O}_{i}(x)$ are generic functions of fields, while $\mathcal{A}_{i}$ are combinations of half-wormhole creation and annihilation operators. 

The effective wormhole action then can be written as 
\begin{widetext}
\begin{equation}
S_{wh}=\int d^{4}x\sqrt{g}\sum_{q}K_{q}e^{-S_{inst}}[(a^{\dagger}_{q}+a_{-q})\mathcal{O}_{-q}(x)+(a^{\dagger}_{-q}+a_{q})\mathcal{O}_{q}(x)]
\end{equation}
\end{widetext}
where the $(a^{\dagger}_{q}+a_{-q})$ describes the creation of a half-wormhole with charge $q$ and the operators satisfy the usual commutation relations. 
In the context of holographic tensor networks, such operators are being represented as the links of those tensors connecting to the central black hole and hence the creation of wormholes is equivalent to legs of internal tensors reaching on the other spacetime region, beyond the wormhole throat. 
These are encoded as the three-indexed tensors in the tensor network. Obviously, such tensors (alternatively three-fields) encode the axion charge and allow the preservation of the axion strong CP solution. Once entanglement is established, we can reconstruct the bulk operators on one side from a certain region $A$ on the boundary provided there exists an operator with support on that region that restores our original operator for any stare belonging to the code Hilbert space. But our operators in this case imply tunnelling transitions and bring the system in a coherent state [9], i.e. 
\begin{equation}
(a^{\dagger}_{-q}+a_{q})\ket{\alpha}=\alpha_{q}e^{i\delta_{q}}\ket{\alpha}
\end{equation}
leading to an effective wormhole action 
\begin{equation}
S_{wh}=\int d^{4}x\sqrt{g}\sum_{q}K_{q}e^{-S_{inst}}\alpha_{q}\mathcal{O}_{S}cos(\frac{q\phi}{f_{a}}+\delta_{q})
\end{equation}
with 
\begin{equation}
\mathcal{O}_{S}=1+aL^{2}R+bL^{4}(\partial_{\mu}\phi)(\partial^{\mu}\phi)+...
\end{equation}
according to [9], but in this case, we interpret the operators as reconstructable from the information available on the other side, and hence the mass shift related to the instanton action is nothing but an effect of the entanglement present between the two sides. 
 Indeed, I showed here that a non-trivial instanton-type contribution to the axion mass can be explained by employing a quantum information language and hence that quantum entanglement implied by the presence of a wormhole has a non-trivial effect on the axion mass. While the non-conservation of the axion charge on one side of the wormhole (or on a half-wormhole) implies the explicit breaking of the symmetry it generates on that side, the axion charge can be considered as overall conserved if one has access to both sides of the wormhole. Of course, this will be restricted in various ways by the non-traversability of the wormhole or, equivalently by the non-signalling properties implied by quantum entanglement. In this article, I show a fundamental new connection between entanglement and mass, by means of holographic tensor networks. Beyond the clear impact on understanding light dark matter candidates, such a connection will shed light on problems related to the experimental discovery of new high temperature superconductors and various types of condensed matter phase transitions.

\end{document}